\documentclass[aps,prb,groupedaddress,twocolumn,10pt]{revtex4-1}

\begin{document}

\newcommand{\be}{\begin{equation}}
\newcommand{\ee}{\end{equation}}
\newcommand{\bea}{\begin{eqnarray}}
\newcommand{\eea}{\end{eqnarray}}

\title{Contrasting string holography to its optical namesake}

\author{D. V. Khveshchenko}

\affiliation{Department of Physics and Astronomy, University of North Carolina, Chapel Hill, NC 27599}

\begin{abstract}
We assess the prospects of using metamaterials for simulating 
various aspects of analogue gravity and holographic correspondence. Albeit requiring 
a careful engineering of the dielectric media, some hallmark features reminiscent of the hypothetical 'generalized holographic conjecture' can be detected by measuring non-local optical field correlations. The possibility of such simulated behavior might also shed light on the true origin of those apparent holography-like phenomena in the 
condensed matter systems with emergent effective metrics which may not, in fact, 
require any references to the string-theoretical holography.
\end{abstract}

\maketitle

%{\it Introduction}

In the past one and a half decade, the 
holographic conjecture that originated from the string theory 
(where it is known as the 'AdS/CFT' correspondence) has crossed 
the inter-disciplinary borders and permeated other fields. 
However, despite the fact that some avid proponents of its broad ('non-AdS/non-CFT') generalizations  
would routinely refer to the latter as 'a well established tool', 
the condensed matter community, by and large, 
continued to hold back from engaging in a substantive discussion 
of the status of this, arguably, the most intriguing paradigm 
shift since the inception of quantum theory.  

Indeed, compared to the original string-theoretical holography  
its {\it ad hoc} applications to the condensed matter and AMO systems \cite{AdS} 
require the most radical (albeit least verified) assumptions,
since such systems are generically neither very strongly coupled, 
nor conformally, Lorentz or even translationally/rotationally invariant 
and lack any supersymmetry or even an ordinary 
gauge symmetry with some rank-$N>1$ (let alone, $N\gg 1$) non-abelian group.

Such striking discrepancies notwithstanding, the available holographic machinery 
makes it almost too (conceptually, if not technically) easy to pursue its applications to the ever expanding list of geometries, thereby favoring computational tractability over physical relevance.  

While some agreement between the holographic predictions and  
certain selectively chosen experimental data (for the most part, pertaining to those situations where the extreme strongly-correlated hydrodynamic regime can indeed be attained) has been claimed,
there is still no consensus on neither the ultimate implications of such circumstantial evidence, nor the general applicability conditions of 
the holographic approach itself. Conspicuously, though, the best quantitative agreement between the results of the holographic and alternative (e.g., Monte Carlo) calculations has been achieved in those cases where the allegedly all-important $N\gg 1$ condition does not seem to play much of a role, as in the $2d$ 
Bose-Hubbard (or quantum $XY$-) model with $N=2$ \cite{MC}.  

Conceivably, such non-compliance of
the hypothetical generalized holography with the symmetry requirements that are viewed as instrumental in its original string-theoretical reincarnation
might suggest that, apart from the common name, the two may not even be related. In that regard, it must be noted that emergent metrics and 
effective gravity-like descriptions are not that uncommon,
such examples ranging from thermodynamics of phase transitions to quantum 
information theory and tensor network states, from 
topological properties of the Bloch states to adiabatic time evolution and Quantum Hall effect, etc.

%{\it Analogue holography}

Nevertheless, even in the current absence of a definitive way to unequivocally ascertain its true status, the holographic concept can still benefit from the possibility of being simulated in various controlled 'analogue' environments.
For one thing, any concrete  physical realization of some apparent holography-like features might contribute towards elucidating the physical phenomena responsible for such behaviors 
without invoking radically new hypothetical principles of nature. 

Recently, one potential implementation of analogue holography 
in the strain-engineered graphene devices was proposed \cite{dvk}. 
In this note, we discuss the prospects of using 
optical metamaterials which have long been 
envisioned as viable candidates for mimicking such effects of general relativity
as event horizons, redshift, black, white, and worm- holes,  
Hawking radiation, dark energy, inflation, multiverse, Big Bang and Rip, 
metric signatures transitions, 'end-of-time', 
and other cosmological scenaria  \cite{ch,smol}.

The above add to the list of such previously explored 
applications of metamaterials as space-time transformation optics, 
negative refraction, sub-diffraction imaging and superlensing, 
cloaking, waveguiding, near-perfect heat absorption/emission, and broadband photonic engineering \cite{lp}.

Specifically, it is a formal mathematical analogy between the Maxwell's equations in the metamaterial media and those of wave propagation in curved space-time that allows one to draw such parallels and use metamaterials as a potential playground for duplicating certain gravitational  phenomena. 
Thus, it puts metamaterials on the already extensive list of the previously proposed experimental implementations of general relativity which includes supersonic fluids, trapped Bose-Einstein condensates of cold atoms and ions, slow light in atomic vapors and nonlinear liquids, 
electromagnetic wavequides, exciton-polariton systems, etc \cite{bh}.

In the metamaterial setups, the locations of  
effective event horizons are determined by the singularities of the permittivity 
$\epsilon_{ij}$ and permeability $\mu_{ij}$ tensors, 
and the early designs of the putative 'Tamm media' involved 
special engineering of both functions \cite{ch}, e.g.:
$
\epsilon^{ij}=\mu^{ij}={\sqrt -{\hat g}}g^{ij}/|g_{\tau\tau}|
$.

The proposed means of creating non-uniform patterns of $\epsilon_{ij}$ and $\mu_{ij}$ include electro-optical modulation and split-ring resonators, respectively. 
while amongst the prospective candidate media are such metal-dielectric pairs as silver and silica, 
$SiC$ and vacuum, as well as plasmonic metals and high index dielectrics $(TiO2, SiN)$.

Still, the task of varying both, $\hat \epsilon$ and $\hat \mu$ - let alone, 
maintaining their equality ('impedance matching') - can hamper any
practical realizations of such metrics. Another impeding factor is the 
invariable value of $g_{\tau\tau}=-1$ which 
limits the choice of the viable metrics to their equivalent ('optical') ones, 
$
\gamma^{ij}=g^{ij}/|g^{\tau\tau}|=\epsilon^{ij}/det{\hat \epsilon}=
\mu^{ij}/det{\hat \mu}
$.  

Another (lower-dimensional, yet potentially more practical) 
approach to the ways in which hyperbolic metamaterials
can be used for desktop simulations of general relativity and cosmology
was put forward in Refs.\cite{smol}.
Their central observation was that for $\mu_{ij}=1$
and under the condition that the photon wave function varies faster than the (non-uniform) 
permittivity $\hat \epsilon$, 
the dispersion relation for extraordinary (TM-polarized) photons 
$
\omega^2={k_z^2/\epsilon_{xy}}+{k^2_{xy}/\epsilon_{zz}}
$
can instead be viewed as that in the empty $3d$ curved space-time with the metric
\be
ds^2=-\epsilon_{xy}dz^2{-\epsilon_{zz}}(dx^2+dy^2)
\ee
In the hyperbolic regime $\epsilon_{xy}>0$ and $\epsilon_{zz}<0$, the component of the momentum $k_z$
behaves as an effective frequency,  while $\omega$ plays the role of mass.

For its most part, the discussion in Refs.\cite{smol} pertained to
the rotationally-invariant uniaxial metamaterial 
configuration, in which case by introducing the cylindrical coordinates and
choosing $z$ and $r$ as the effective time $\tau$ and natural 
'radial' holographic variable, respectively, one arrives at the metric
$ 
g_{\tau\tau}=-\epsilon_{xy}, ~~~g_{rr}=g_{\phi\phi}/r^2=-\epsilon_{zz}
$.

The attainable metrics would then be described by the diagonal dielectric tensor 
of a two-component system with the permittivities $\epsilon_{m}<0$ and $\epsilon_{d}>0$
('metal' and 'dielectric', respectively). The dielectric properties of such a combination
can be evaluated by using the Maxwell-Garnett formula 
\bea
\epsilon^{(u)}_{zz}(r)={\epsilon_1 n+\epsilon_2(1-n)}\nonumber\\
\epsilon^{(u)}_{xy}(r)=\epsilon_2{\epsilon_1 (1+n)+\epsilon_2(1-n)\over \epsilon_2 (1+n)+\epsilon_1(1-n)}
\eea 
where the structure factor $n(r)$ represents the local (radially-dependent) fraction
of the metallic component. 
The analysis of Eq.(2) shows that the viable metrics can possess 
either a pole or zero in $\epsilon^{(u)}_{xy}(r)$ but only a zero in $\epsilon^{(u)}_{zz}(r)$, although those would  
generally occur at different values of the function $n(r)$ and, therefore, different radial distances. 

Contrary to the assertion of Refs.\cite{smol}, though,  
the setup in question would be unsuitable for mimicking any $2+1$-dimensional black 
hole type of geometry, as the latter requires the simultaneous presence of a zero in $g_{\tau\tau}$ and 
a pole in $g_{rr}$  at some putative horizon $r_h$  (i.e., $g_{\tau\tau}\sim 1/g_{rr}\sim r-r_h$),   
whereas $g_{\phi\phi}$ should develop neither.
Likewise, modelling an optical counterpart of the Schwarzschild-like metric would require
 $\gamma_{rr}\sim\gamma^2_{\phi\phi}\sim (r-r_h)^2$ and, therefore, is out of reach as well.

Instead, one finds that $g_{\tau\tau}$ containing the would-be emblackening factor 
features either a zero at $n_1=(\epsilon_m+\epsilon_d)/(\epsilon_d-\epsilon_m)$
for $|\epsilon_m|<\epsilon_d$ or a pole at $-n_1$ for $|\epsilon_m|>\epsilon_d$.  
In turn, the components $g_{rr}=g_{\phi\phi}/r^2$ can develop a zero at 
$n_2=\epsilon_d/(\epsilon_d-\epsilon_m)$. 
Moreover, for $\epsilon_m=-2\epsilon_d$ 
the zero in $g_{rr}\sim g_{\phi\phi}$ and pole in $g_{\tau\tau}$ merge
together at $n_2=-n_1=1/3$.

In this case, choosing the density 
profile in the near-boundary (i.e., large radii) regime as   
$
n(r)=1/3+c(R/r)^{2\alpha}
$, 
where $c\ll 1$,  $x=R\phi$ and  $u=R/r$   
is the customary holographic 'inverse radial' variable,  
produces a metric of the general type  
\be
ds^2={d\tau^2\over u^{2\alpha}}+R^2{du^2\over u^{2\beta}}+{dx^2\over u^{2\gamma}}
\ee 
with the exponents related as follows: $\beta=2-\alpha,~~\gamma=1-\alpha$.  

For arbitrary values of $\alpha,\beta$, and $\gamma\neq 0$
the metric (3) can be conformally transformed to that of the 
hyperscaling violation (HV) variety \cite{hv}
\be
ds^2=u^{2\theta/d}({d\tau^2\over u^{2\zeta}}+{L^2du^2+d{\bf x}^2\over u^{2}})
\ee  
where $\bf x$ is the $d$-dimensional spatial vector (henceforth, $d=1$),
$L\sim R$ is the overall length scale (the size of the $AdS$ space for $\alpha=1$), while 
the dynamical exponent $\zeta$ and the HV parameter $\theta$ are given by the expressions   
\be
\zeta={1-\beta+\alpha\over 1-\beta+\gamma},~~~\theta={1-\beta\over 1-\beta+\gamma}
\ee
Given the above relations between the exponents
in Eq.(4), however, the only attainable geometry of the HV-type appears to be that of
the extremal limit where $\theta, z\to\infty$, while
their ratio $\theta/\zeta =(\alpha-1)/(2\alpha-1)$ remains finite.

In contrast, for the special values $\gamma=0,~~\alpha=\beta=1$ the metric (4) 
conforms to the zero-temperature Euclidean $AdS_2\times R$ background
(a lower-dimensional counterpart of the space $AdS_2\times R^2$ which provides the arena 
for the popular 'semi-local' holographic scenario \cite{semi}).

As an alternative, one can exploit the zero of $g_{\tau\tau}$ 
which develops for $|\epsilon_m|<\epsilon_d$ and which is not  
accompanied by the zero in $g_{rr}$. In this case, the density profile $n(r)=n_1+c(R/r)^2$ 
gives rise to the geometry resembling the zero-temperature
limit of the Rindler's metric, $ds^2=(r/R)^2d\tau^2+dr^2+(r/R)^2dx^2$, 
although the actual horizon at $r=0$ would be unattainable within its regime of validity.

Also, by tuning the density to the value
$n_2\neq \pm n_1$ and making use of the 
zero in $g_{rr}\sim g_{\phi\phi}$ one could achieve the 'end-of-time' situation \cite{smol}, 
albeit only in the case of a boundary which is orthogonal to the effective time direction.

Lastly, for a generic choice of $\epsilon_{m,d}$ the pole and zeros in Eq.(1)   
are all separated and the corresponding geometry bears 
no resemblance to any physically relevant one 
(this includes the original proposal of Refs.\cite{smol} which makes use of the sole pole in $g_{\tau\tau}$).

In the complementary, layered, metamaterial configuration,   
the composition rules for the dielectric tensor 
of a two-component hybrid system read
\bea
\epsilon^{(l)}_{zz}(z)={\epsilon_1\epsilon_2\over \epsilon_1(1-n)+\epsilon_2n}\nonumber\\
\epsilon^{(l)}_{xy}(z)={\epsilon_1 n+\epsilon_2(1-n)}
\eea 
thus revealing a potentail zero in $\epsilon^{(l)}_{xy}(z)$ and a pole in $\epsilon^{(l)}_{zz}(z)$ 
at, generally, two different values of $n(z)$.
The two coincide at $\epsilon_m=-\epsilon_d$, though, in which case by choosing a power-law 
density profile near the $z=0$ boundary, 
$
n(z)=1/2+c(z/L)^{2\alpha}
$,
and designating $u=z$ as the holographic variable, one arrives at the HV metric (4) with
$\alpha=-\beta=\gamma$. The Euclidean time $\tau$ and the $1d$ spatial coordinate 
$x$ can be chosen arbitrarily within the rotationally invariant $xy$ plane,
while the length scale $L$ is now set by
the span of the region supporting the above algebraic density profile in the $z$-direction. 

Using Eq.(5), one then finds 
$
\zeta=1,~~\theta={(1+\alpha)/(1+2\alpha)}
$.
Although for $\alpha=-1$ the metric (4) may seem to be that of the Euclidean $AdS_3$, such a behavior would be limited to the small, rather than asymptotically large, values of $u$. 

Besides, the effective metrics corresponding to neither the uniaxial, nor layered
configurations could incorporate the thermal emblackening factor replacing the constant $1/c$
and vanishing at some $u_h\sim 1/T$, as it would result in the unwanted 
divergence of $g_{\phi\phi}$ or vanishing of $g_{xx}$, respectively. 
It is also worth mentioning that in all the above situations 
both, $\epsilon_{zz}$ and $\epsilon_{xy}$, have the same 
sign (for either sign of $c$), and so the metamaterial as a whole remains in 
the (anisotropic) metallic or insulating regime. 

%{\it Boundary correlation function}

In the semiclassical limit $\omega L>>1$ (here $\omega$
plays the role of mass which is important for justifying the use of this approximation), the propagator of a massive $3d$ field in a radially-symmetric bulk metric is governed by the classical 
Euclidean action 
\be
S(\tau, x)=L\omega\int du{\sqrt {g_{uu}+g_{\tau\tau}({d\tau\over du})^2+
g_{xx}({d x\over du})^2}} 
\ee
whose extremal paths correspond to the geodesic trajectories.
In particular, by choosing their endpoints to lie on the boundary one accesses   
the geodesics which dive into the bulk, thus exploring its geometry, and saturate the
semiclassical boundary propagator $G_{\omega}(\tau, x)\sim\exp[-S(\tau, x)]
$.
 
After having been evaluated upon such an extremal trajectory, Eq.(7) takes the form
\be
S(\tau, x)=L\omega^2\int^{u_t}_{u_0} du{\sqrt {g_{uu}\over r(u)}}
\ee
where $r(u)=\omega^2-k^2_x/g_{xx}(u)-k^2_\tau/g_{\tau\tau}(u)$ 
is a function of the conjugate momenta 
$k_x=\delta S/\delta (dx/du)$ and
$k_\tau=\delta S/\delta (d\tau/du)$.
The latter, in turn, are determined by the equations
\be
\tau=Lk_\tau
\int^{u_t}_{u_0} {du\over g_{\tau\tau}}{\sqrt {g_{uu}\over r(u)}},~~~
x=Lk_x
\int^{u_t}_{u_0} {du\over g_{xx}}{\sqrt {g_{uu}\over r(u)}}
\ee
where $u_0\sim 1/L$ and the turning point satisfies $r(u_t)=0$. 

For the sake of generality, we compute Eq.(8)  
in the general case of the power-law permittivities 
,  
$
\epsilon_{zz}(u)\sim (L/u)^{2\alpha},~~\epsilon_{xy}(u)\sim (u/L)^{2\delta}
$,
thus obtaining 
$
S(\tau, x)\sim |{\bf x}|^{(1+\delta)/(1+\alpha+\delta)}
$
as long as $\alpha>0$ and $\delta>-1$. This condition guarantees that the 
integrals in Eqs.(8,9) are dominated by the turning point 
$u_t=(\omega/{\sqrt {k_\tau^2+k_x^2}})^{1/\alpha}$
which moves deeper into the bulk with the increasing Euclidean 
interval on the boundary, $|{\bf x}|={\sqrt {\tau^2+x^2}}$.
The semiclassical condition 
$S({\bf x})\gg 1$, alongside the aforementioned applicability of Eq.(1),   
dictate that  
$c^{1-\zeta/2\theta}(L\omega)^{-\zeta/\theta}\ll x/L\ll 1$, 
which conditions are compatible for $c\ll 1$ and $L\omega\gg 1$.

In either of the above metamaterial configurations, though, the function $n(u)$ 
yields just one independent exponent, hence $\alpha=\delta$.
In the layered case, one then obtains a stretched exponential asymptotic of the propagator 
\be
G_{\omega}({\bf x})\sim\exp[-{\sqrt c}L\omega |{\bf x}/cL|^{\theta/\zeta}]
\ee
where the HV parameter-to-dynamical exponent ratio
falls into the range $1/2<\theta/\zeta=(1+\alpha)/(1+2\alpha)<1$. 

In contrast, for the uniaxial radial-dependent density distribution with $\alpha<1$ 
the integrals in Eqs.(8,9) are dominated by the lower bound $u_{0}\sim 1/R$ and, therefore, 
become insensitive to the position of the turning point $u_t$, thus rendering 
$S(\tau,x)\sim R^{1-\alpha}$. This independence 
of the extremal action of its arguments 
indicates that the logarithm of the boundary propagator varies slower than any
fractional exponent of either $\tau$ or $x$. 
In fact, the decay of $G_{\omega}(\tau, x)$ appears to be algebraic and governed by the non-exponential prefactor
which is beyond the leading order of the semiclassical approximation. 

At $\alpha=1$ one finds $S(\tau,0)\sim\ln\tau$ and $S(0,x)\sim x$,
which dependencies are consistent with the 'semi-local' regime 
characterized by the long-ranged (power-law) temporal, yet only 
short-ranged (exponential) spatial, decay 
of the propagator $G_{\omega}(\tau, x)$ \cite{semi}.

Lastly, for $\alpha>1$ one again finds a stretched exponential   
time dependence, $S(\tau,0)\sim \tau^{\theta/\zeta}$, 
where $0<{\theta/\zeta}=(\alpha-1)/(2\alpha-1)<1/2$. However,  
the spatial correlations decay extremely fast, thus reaching the truly localized 
limit, which is fully consistent with $\zeta\to\infty$. 

Incidentally, the behavior (10) is akin to that found in, e.g., the studies of relaxation in 
disordered media whose field-theoretical description implies a strongly coupled 
nature of the putative $2d$ boundary theory.
To that end, it can be reproduced by choosing the standard exponential construction
of the boundary operators $\psi({\bf x})\sim\exp[i\phi({\bf x})]$ in terms of a $2d$ scalar bosonic field 
whose Lorentz-invariant action is governed by the quadratic part 
\be
S_{boundary}={1\over 2\nu}\int d^2{\bf k} {\bf k}^{2+\theta/\zeta}|\phi_{\bf k}|^2
\ee 
where $\nu\sim L^{1-\theta/\zeta}\omega$.
Notably, the (anomalous) dimension $[\phi]=-\theta/2\zeta$ is negative, thereby resulting in the correlation function $<|\phi_{\bf k}|^2>$ which diverges faster than $1/{\bf k}^2$ at small $\bf k$. 

Incidentally, the effective Gaussian action similar to Eq.(10) emerges in the 
theory of thermally fluctuating membranes governed by the elastic free energy \cite{lr} 
\be
F=\int d^d{\bf x}[{\kappa\over 2}(\nabla^2h)^2+\mu v_{\alpha\beta}^2+{\lambda\over 2}v_{\alpha\alpha}^2]
\ee 
where $v_{\alpha\beta}=\partial_{\alpha}\xi_{\beta}+\partial_{\beta}\xi_{\alpha}+
\partial_{\alpha}h\partial_{\beta}h$ is the strain tensor expressed 
in terms of the in-plane $\xi_\alpha({\bf x})$ and out-of-plane $h({\bf x})$ displacements. 

Upon integrating out the former, the effective self-energy 
of the latter is generated, $\Delta F\sim \int d^d{\bf k} {\bf k}^{4-\eta}|h_{\bf k}|^2$,
thus overpowering the first term in Eq.(12) at small $\bf k$.
The anomalous exponent $\eta$ characterizes the behavior of the fluctuation-renormalized 
bending rigidity $\kappa\sim 1/{\bf k}^{\eta}$, taking the value $\eta_c\sim 1$
at the crumpling transition of a physical ($2d$) membrane 
embedded in the $3d$ space. 

As compared to the standard (marginally non-Fermi liquid) Luttinger regime where the vertex operators
are given by exponentials of the ordinary $1d$ phonon field with a linear dispersion,
the ultrasoft out-of-plane ('flexor') 
phonon modes with the free energy (11) result in the expressly non-Fermi liquid behavior.
Also, in contrast to the holographic semi-local regime where the operator dimensions 
appear to be featureless continuous functions of the momentum \cite{semi}, the pertinent 
exponent $\theta/\zeta$ is a robust function of the given metric (4). 

Thus, by properly adjusting the density profile $n(z)$ of the metallic component 
one could, in principle, use the proposed setup for mimicking the correlation functions 
of a certain family of the non-trivial $2d$ boundary theories.
Notably, though, establishing such a limited correspondence does not require
any novel principles of nature, including the sophisticated string-theoretical 
constructions. However, if observed in, e.g., numerical studies of the theory (12) such a matching behavior
could be (misleadingly) argued to support the generalized holographic conjecture. 

In the metamaterial implementations,
the propagator (10) describes static boundary correlations of the electromagnetic field and, as such, should
be contrasted against its counterpart in a medium with the isotropic (negative) dielectric constant 
$
<E_{\omega}({\bf x})E^{*}_{-\omega}(0)> \sim \exp(-\omega |{\bf x}|)
$.
Experimentally, such correlations can be studied by analyzing the 
statistics of the spatial distribution of a monochromatic optical field at the metamaterial's boundary.
Conceivably, the actual measurement should employ such techniques as holographic and 
speckle interferometry \cite{hi},
by which a comparison can be made between the light that travels thorough the bulk and the outside reference beam. 
In particular, by measuring the local correlations, $S_2(\omega)=<|E_{\omega}(0)|^2>\sim G_{\omega}(0)$,  
one can access noise power spectrum (although the semiclassical 
result (10) would be inapplicable for ${\bf x}\to 0$). 

Also, calculating the higher-order correlations, such as noise of noise $S_4(\omega)$, one can ascertain the (non-)Gaussianity of the fluctuations and applicability of the quadratic effective action (11). 
Moreover, the field distribution might be accompanied by a correlated 
pattern of currents associated with the local plasmon-polariton modes at the
conducting  metamaterial's boundary, thus allowing 
for alternate detection techniques based on the measurements of Johnson-Nyquist current noise.  

As to the practical realizations, in Refs.\cite{smol} 
the uniaxial devices were constructed with the use of ferromagnetic cobalt (or silver/gold)
nanoparticles floating in a dielectric liquid, such as kerosin, 
which align themselves in filaments in the presence of a magnetic field in the $z$-direction.
Also, the polymer (PMMA) stripes deposited on gold films have been utilized \cite{smol}.
Regarding the layered design, it can be manufactured from alternating 
metallic and dielectric layers of variable width. 
One such proposal involves $n^{+}$-doped $InGaAs$ for the metallic 
and $AlGaAs/GaAs$ for the dielectric components, respectively  \cite{1402.4475}. 

To arrive at one of the pole-zero matching points one can 
utilize the frequency dependence of $\epsilon_{m,d}$.
For example, in the devices used in Refs.\cite{smol} the zero of $\epsilon_{xy}$ becomes attainable at the long wavelength infrared frequencies.

%{\it Summary}

In conclusion, 
we discussed the prospects of using metamaterial devices for mimicking 
certain holographic correspondence-like phenomena. 
To that end, we identified the class of metrics that, in principle, can be reproduced in the two  
popular (uniaxial and layered) architectures and evaluated the correlation function of  
monochromatic electromagnetic fluctuations at the metamaterial's boundary.

The latter function was also shown to be obtainable as the correlator of vertex operators in the theory of a 
strongly self-interacting $2d$ bosonic field, akin to that describing the thermodynamics of
a fluctuating elastic membrane. This observation opens up the possibility of, both, simulating 
certain non-trivial boundary theories with custom-tailored
metamaterial media as well as gaining a better insight into the potential origin of
those apparent holography-like properties of condensed matter systems 
which may not, in fact, require any strained references to (yet, can be misinterpreted as an evidence for) 
some far-fetched generalizations of the 'bona fide' holographic principle of string 
theory.

The author acknowledges the Aspen Center for Physics funded 
by the grant NSF under Grant 1066293 for its hospitality and workshop 
participation support.

\end{document}